\documentstyle[epsfig,cite]{mn}

\begin{document}

\title[Outburst of XTE J1859+226]
    {Initial low/hard state, multiple jet ejections and X-ray/radio correlations during the outburst of XTE J1859+226}
\author[Brocksopp et al.]
    {C.~Brocksopp$^1$\thanks{email: cb@astro.livjm.ac.uk}, R.P.~Fender$^2$, M.~McCollough$^3$, G.G.~Pooley$^4$, M.P.~Rupen$^5$,
\newauthor
R.M.~Hjellming$^5$, C.J.~de~la~Force$^6$, R.E.~Spencer$^6$, T.W.B.~Muxlow$^6$, 
\newauthor
S.T.~Garrington$^6$, S.~Trushkin$^7$\\
\\
$^1$Astrophysical Research Institute, Liverpool John Moores University, 12 Quays House, Egerton Wharf, Birkenhead CH41 1LD\\
$^2$Astronomical Institute ``Anton Pannekoek'' and Center for High-Energy Astro
physics, University of Amsterdam,\\
\hspace{0.5cm}Kruislaan 403, 1098 SJ Amsterdam, The Netherlands\\
$^3$Universities Space Research Association, Huntsville, AL 35812, USA\\
$^4$Mullard Radio Astronomy Observatory, Cavendish Laboratory, Madingley Road, Cambridge CB3 0HE\\
$^5$National Radio Astronomy Observatory, P.O. Box O, Socorro, NM 87801, USA\\
$^6$Jodrell Bank Observatory, Macclesfield, Cheshire, SK11 9DL\\
$^7$Special Astrophysical Observatory, RAS, Nizhnij Arkhyz, Karachaevo-Cherkassia 369167, Russia\\}
\date{Accepted ??. Received ??}
\pagerange{\pageref{firstpage}--\pageref{lastpage}}
\pubyear{??}
\maketitle

\begin{abstract}

We have studied the 1999 soft X-ray transient outburst of XTE J1859+226 at radio and X-ray wavelengths. The event was characterised by strong variability in the disc, corona and jet -- in particular, a number of radio flares (ejections) took place and seemed well-correlated with hard X-ray events. Apparently unusual for the {\em canonical} `soft' X-ray transient, there was an initial period of low/hard state behaviour during the rise from quiescence but {\em prior} to the peak of the main outburst -- we show that not only could this initial low/hard state be an ubiquitous feature of soft X-ray transient outbursts but that it could also be extremely important in our study of outburst mechanisms.

\end{abstract}


\section{Introduction}

\begin{figure*}
\begin{center}
\leavevmode  
\psfig{file=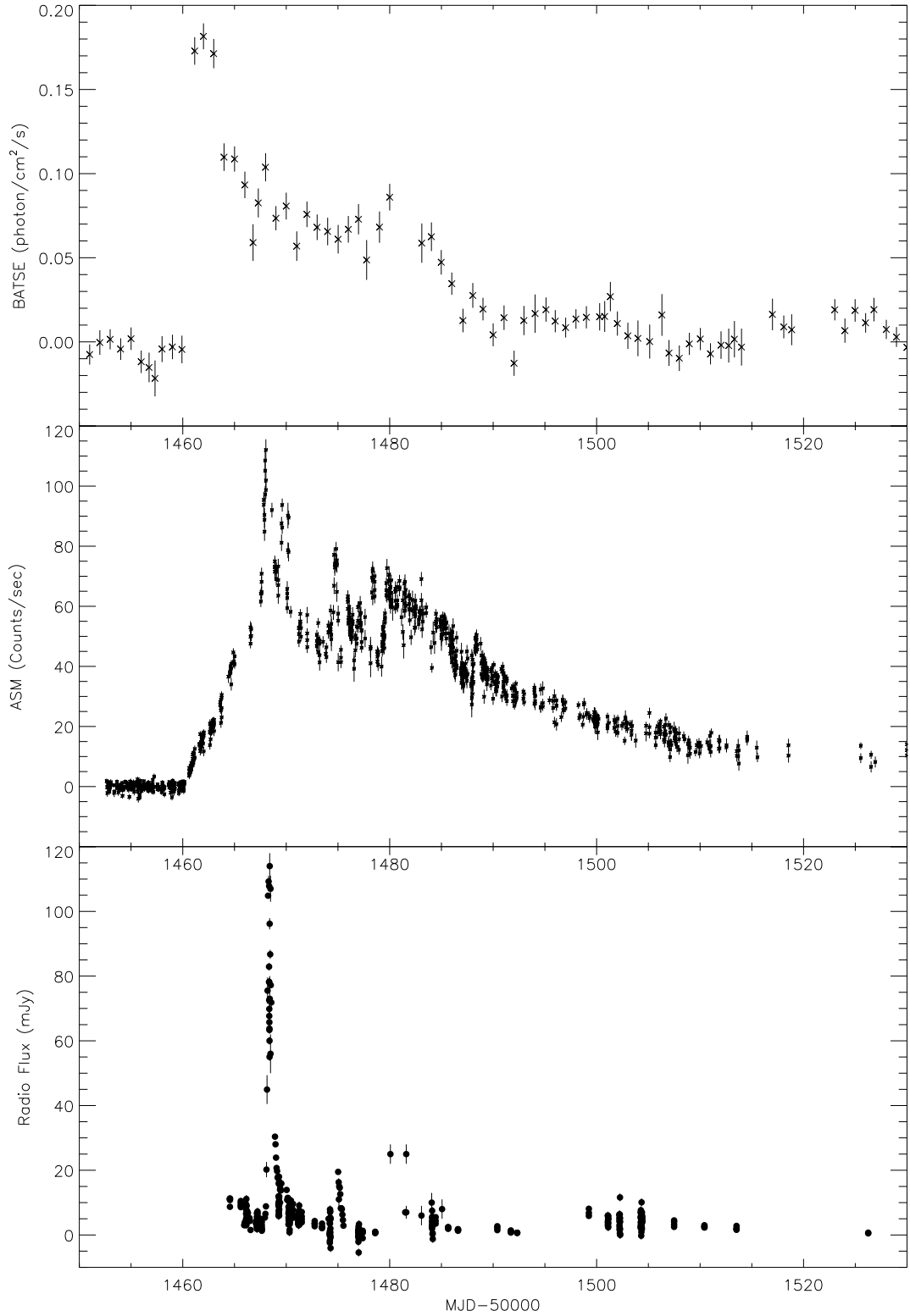,height=22cm,angle=0}  
\caption{Radio lightcurve from the VLA, MERLIN, the Ryle Telescope, the RATAN Telescope and a few additional points from the GBI plotted with the hard and soft X-ray data from {\sl CGRO}/BATSE and {\sl RXTE}/ASM.}
\label{lightcurve}
\end{center}
\end{figure*}

\begin{figure}
\begin{center}
\leavevmode  
\psfig{file=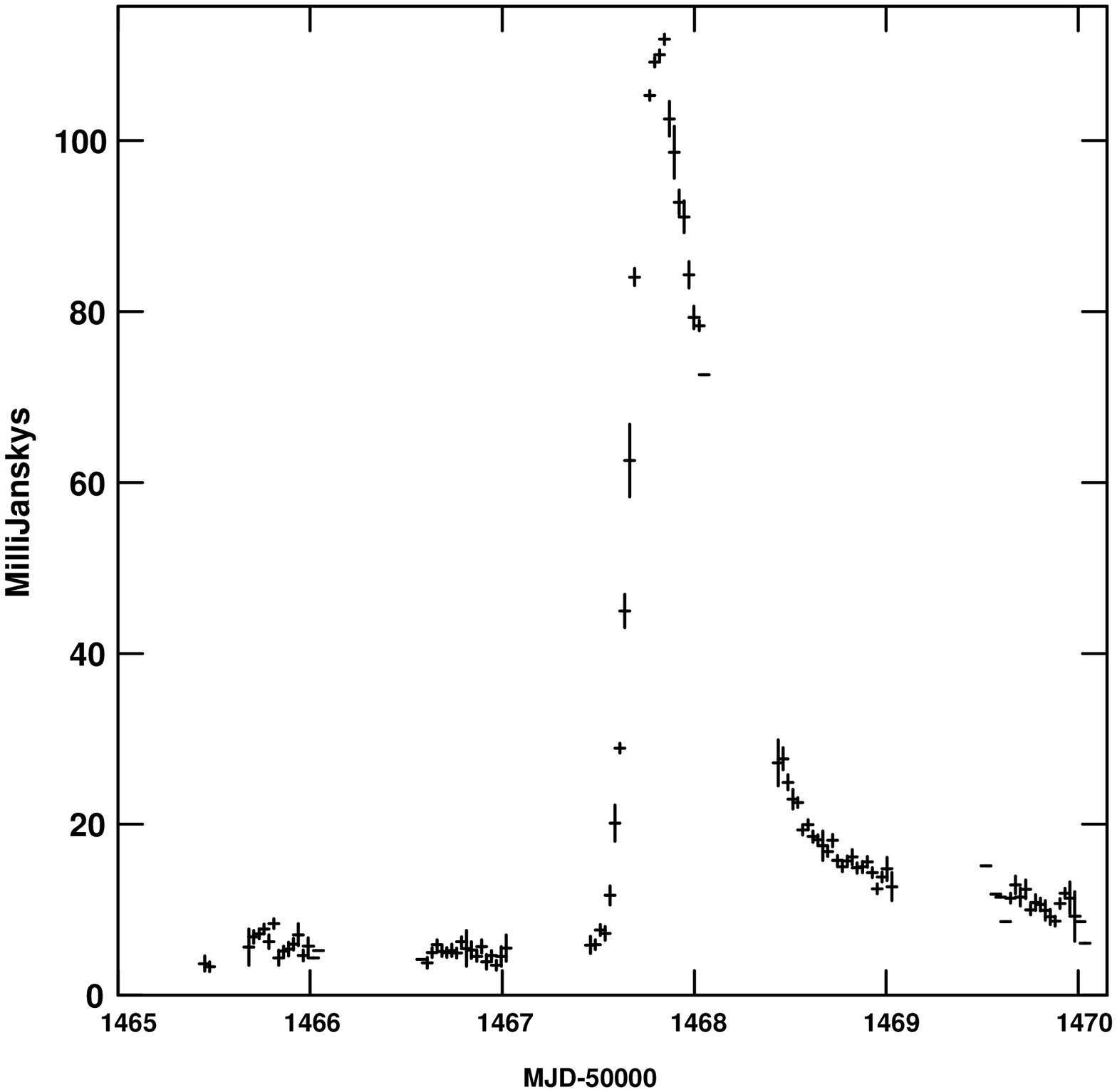,height=8cm, angle=0}  
\caption{The 1.66 GHz lightcurve of MERLIN, showing the excellent coverge of the first and major flare.}
\label{merlin}
\end{center}
\end{figure}

X-ray transients are a group of X-ray binaries which exhibit dramatic outbursts, generally at all wavelengths, due to changes in mass accretion rate. A `typical' soft X-ray transient event (or `X-ray nova')  will show a fast rise to its maximum luminosity, accompanied by X-ray spectral softening, and reach the high or sometimes the very high state; this is then followed by a slow exponential decay back to quiescence (see M\'endez et al. 1998). More recently fewer transients have appeared to behave in this typical manner, almost certainly due (in part?) to our improved observing coverage. Some show a period of hard state behaviour before reaching the full outburst (e.g. XTE J1550$-$564 Wilson \& Done 2001); others do not progress beyond the low/hard state at all (the hard state transients -- see e.g. Brocksopp et al. 2001 and references within). Note that the terms `low' and `high' states do not necessarily refer to the flux of a source but instead indicate the relative contributions of hard and soft X-ray flux -- see van der Klis (1995) and Nowak (1995) for full details of spectral state classification.

XTE J1859+226 was discovered by the All Sky Monitor onboard the Rossi X-ray Timing Explorer satellite on 1999 October 9 (MJD 51460; Wood et al. 1999). On its initial detection the soft (2--12 keV) X-ray flux was $\sim$ 160 mCrab and rising at $\sim$ 6 mCrab/hour. Follow-up PCA observations revealed a hard power law energy spectrum and a power spectrum with a 0.45 Hz quasi-periodic oscillation (QPO), plus harmonics (Markwardt et al. 1999). Hard X-ray observations by BATSE (of the Compton Gamma Ray Observatory) confirmed the hard spectrum up to 200 keV and showed that the hard X-ray flux peaked while the soft X-ray source was still rising (McCollough \& Wilson 1999). We note that this `initial low/hard state' is not thought to be a common property of a `canonical' soft X-ray transient outburst, although we show in Section 4 that it has indeed been observed many time previously. These X-ray properties suggested a likely black hole X-ray binary nature for the system.

Despite the intially hard spectrum, suggesting that XTE J1859+226 may be an addition to the list  of hard state X-ray transients (Brocksopp et al. 2001; Brocksopp, Bandyopadhyay, Fender 2001, in prep.), as the hard X-ray flux decreased the soft X-ray source entered a series of flares; the source softened and reached a peak of $\sim$ 1.5 Crab (2--12 keV) about eight days after the initial detection (Focke et al. 2000). PCA observations durin g this flaring period again showed the presence of QPOs, this time at 6--7 Hz and a second at 82--187 Hz (Cui et al. 2000).

The radio counterpart was discovered at a flux density of $\sim$ 10 mJy on 1999 October 11 (MJD 51462) at the Ryle Telescope and VLA (Pooley \& Hjellming 1999). The optical counterpart was also discovered a day later with an $R$-band magnitude of $\sim$ 15.1 (Garnavich et al. 1999). Spectroscopy of this source revealed a strong blue continuum with weak emission lines at H$\alpha$, H$\beta$, He\,{\sc ii}\,$\lambda 4686$ and C\,{\sc iii}/N\,{\sc iii}\,$\lambda 4640--4650$ (Garnavich et al. 1999, Wagner et al. 1999). Studies of an interstellar absorption feature suggested a low value ($E(B-V)\sim 0.58$) for the interstellar extinction (Wagner et al. 1999, Hynes et al. 1999).

Following decay of the outburst, attempts have been made to search the optical photometry for the orbital period of the system and a potential modulation at 9.15 hours discovered (Garnavich \& Quinn 2000); this was confirmed by Sanchez-Fernandez et al. (2000). More recently a mass function of 7.4 M$_{\odot}$ has been determined, thus confirming this suggestion that the compact object is a black hole (Filipenko \& Chornock 2001). A QPO of 0.76 mHz was also detected in optical time series data (Chaty et al. 2001).

X-ray and optical observations have classified XTE J1859+226 as a `typical X-ray nova' with a fast rise and exponential decay (Chaty et al. 2001). Comparisons with various other X-ray transients have been drawn -- it entered an optical `mini-outburst' 235 days after the initial maximum, similar to GRO J0422+32 (a {\em hard} state transient) and its optical behaviour has also been likened to that of A0620$-$00 and GRS 1124$-$68 (Charles et al. 2000). Alternatively, the X-ray behaviour is reminiscent of XTE J1550$-$564 (Cui et al. 2000).

In this paper we study the radio and X-ray observations of XTE J1859$-$226. The observations are outlined in Section 2 and the results presented in Section 3. Finally we compare certain features of XTE J1859+226 with those of other sources in Section 4 and discuss the implications of these results in Section 5, before drawing our conclusions in Section 6.

\begin{figure}
\begin{center}
\leavevmode  
\psfig{file=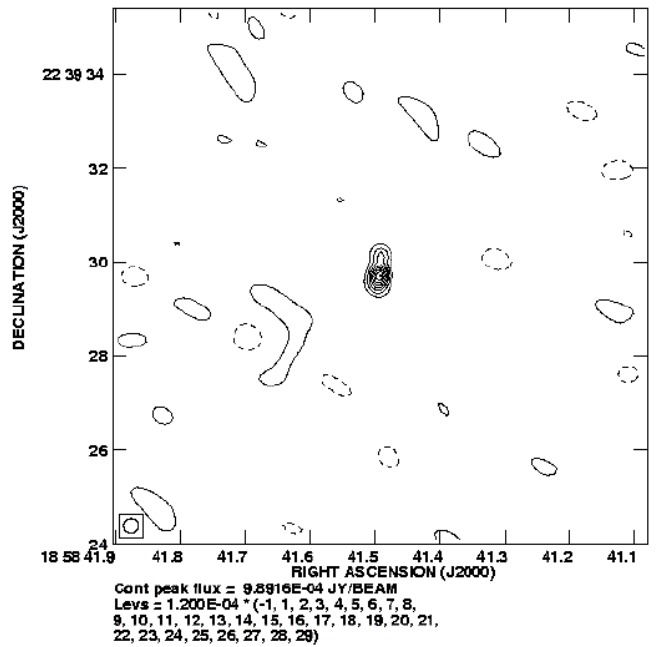,height=9cm}  
\caption{8.64 GHz map of XTE J1859+226 taken by the VLA on 1999 December 1. It appears to be resolved, thus confirming our predictions throughout this paper that XTE J1859+226 is a jet source. Further analysis of the radio maps will be presented in a future paper.}
\label{map}
\end{center}
\end{figure}

\section{Observations}

\begin{figure*}
\begin{center}
\leavevmode  
\psfig{file=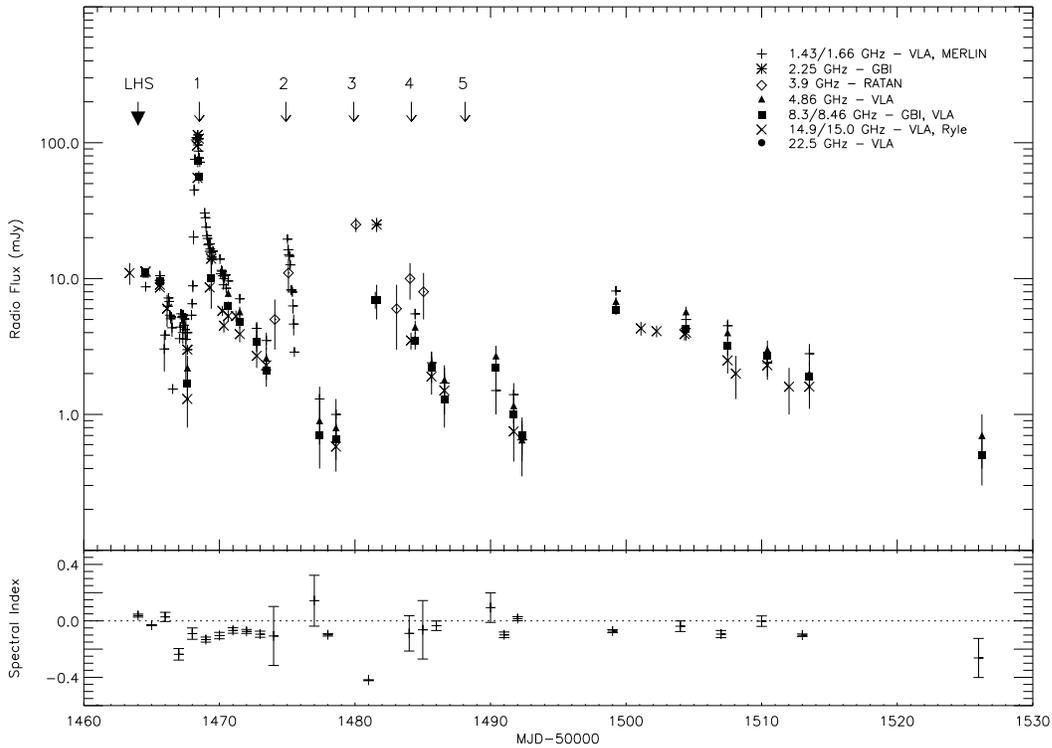,height=14cm,angle=90}  
\caption{The radio lightcurves plotted on an expanded scale and using different symbols to indicate different frequencies. Arrows indicate the estimated time of each ejection (and also the initial low/hard state, denoted by LHS) and correspond to the circled points in Figs.~\ref{hardness} and \ref{colour}. The bottom panel shows the spectral index at each epoch where there were sufficent data for it to be measured (see Table~\ref{alpha}).}
\label{radio-lightcurve}
\end{center}
\end{figure*}

Details of each telescope/satellite and their observations are outlined below. All data points obtained are shown in Fig.~\ref{lightcurve}.

\subsection{Radio}

\subsubsection{VLA}
 
Data were obtained by the Very Large Array in New Mexico at 1.43, 4.86, 8.46, 14.9 and 22.5 GHz, at a total of 52 epochs between 1999 October 13 and 1999 December 13 (MJD 51464-525).  In each case 100 MHz was observed in each of two orthogonal polarizations.  The flux scale is that derived by Perley \& Taylor from VLA observations in 1995.2 (cf. the VLA Calibrator Manual).  Phases were calibrated using two nearby standard VLA calibrators, 1850+284 and 1925+211. The data were taken in a variety of configurations, but mostly in the relatively extended BnA and B configurations.  Further observations from 2000 October 1 through 2001 February 6 (MJD 51818-946) gave $3\sigma$ upper limits between 0.18 and 0.45 mJy/beam at 8.46 GHz.

\subsubsection{MERLIN}

Seven epochs of MERLIN (Multi-Element Radio Linked Interferometer, UK) data were obtained at 1.658 GHz (bandwidth 15 MHz) between 1999 October 13 and 1999 October 22 (MJD 51464-473). Six antennas were used -- Defford, Cambridge, Knockin, Darnhall, Mk2 and Tabley. The flux calibrator was 3C286, the point source calibrator was OQ208 and the phase calibrator 1851+236. We plot the early MERLIN data in Fig.~\ref{merlin} in order to show its very good coverage of the first and major flare; we note that it is rare to obtain observations of the rise of a flare.

\subsubsection{Ryle}
Further 15 GHz radio monitoring was obtained from the Ryle Telescope of the Mullard Radio Astronomy Observatory, Cambridge, UK between 1999 October 11 and 1999 November 21 (MJD 51462-503). Further details of the observing technique may be found in Pooley \& Fender (1997).

\subsubsection{RATAN}

XTE J1859+226 was observed with the north sector of the RATAN 600 metre telescope at 3.9 GHz and with a bandwidth of 700 MHz. Simultaneous observations of calibrator NVSSJ190116+223844 were also used, along with quasar 3C286 as flux calibrator. Data were obtained at six epochs between 1999 October 22 and 1999 November 2 (MJD 51473-484). 

\subsubsection{Green Bank Interferometer}

Additional radio points at 2.25 and 8.3 GHz have been obtained from the public database of the Green Bank Interferometer. Details of the observations and their calibration can be found in Waltman et al. (1994).

\subsection{Soft X-ray}

We have obtained soft (2--12 keV) X-ray data from the {\it RXTE}/ASM -- we use the public archive data from the web (http://xte.mit.edu). A detailed description of the ASM, including calibration and reduction is published in Levine et al. (1996).

\subsection{Hard X-ray}

The BATSE experiment onboard {\it CGRO} (Fishman et al. 1989) was used to monitor the hard X-ray emission. The BATSE Large Area Detectors (LADs) can monitor the whole sky almost continuously in the energy range of 20 keV -- 2 MeV with a typical daily 3$\sigma$ sensitivity of better than 100 mCrab.  Detector counting rates with a timing resolution of 2.048 seconds are used for our data analysis. To produce the XTE J1859+226 light curve, single step occultation data were taken using a standard Earth occultation analysis technique used for monitoring hard X-ray sources (Harmon et al. 1992a). Interference from known bright sources were removed. A spectral analysis of the BATSE data indicated that the data (average of the first three days of the outburst) were well-fit by a power law with a spectral index of $-$2.6. The single occultation step data were then fit with a power law with this index to determine daily flux measurements in the 20--100 keV band. We note that determining the fluxes from a variable spectral index did not alter the lightcurve significantly.

\section{Results}

The radio counterpart to XTE J1859+226 was imaged by the VLA and MERLIN at various frequencies. While both sets of images appeared asymmetrically extended in an approximately northwards direction, the resolution at most epochs was insufficient to confirm the presence of a jet. However, VLA observations on 1999 December 1 (MJD 51513) at 14.9 and 8.46 GHz did appear to be resolved (Fig.~\ref{map}) -- this is to be investigated in a future paper. We now use further analysis in the rest of this section to suggest that a jet was indeed present, regardless of whether or not our observations were able to resolve it.

The radio and X-ray data are plotted in Fig.~\ref{lightcurve}. It is interesting to see that initially the X-rays do not show the behaviour of a typical soft X-ray transient -- instead of the soft X-rays rising rapidly they brighten gradually and it is the hard X-rays that increase quickly into a luminous flare. Neither is there a `classical' smooth exponential decay -- as the hard X-rays decline, the soft X-rays and radio emission flare simultaneously before entering a series of smaller, apparently correlated flares, some of which also take place in the hard X-rays.

With such a high level of variability it is perhaps suprising that the accompanying behaviour in the optical is just a smooth decline -- indeed, it is not until the infrared ($K$ band) that there is any hint of flaring variability and in this case there is insufficient data to be certain (Chaty et al. 2001).

\subsection{Radio}

\begin{table}
\center
\caption{The spectral index calculated for each epoch of our radio observations. The fourth column indicates the number of frequencies which available at each epoch from which to fit a power law and calculate $\alpha$.}
\label{alpha}
\begin{tabular}{cccc}
\hline
MJD$-$50000&$\alpha$&error&No. of points\\
\hline
1464 &   0.037 & 0.011 & 5\\
1465 &  -0.030 & 0.004 & 5\\
1466 &   0.028 & 0.033 & 2\\
1467 &  -0.237 & 0.041 & 7\\
1468 &  -0.090 & 0.040 & 4\\
1469 &  -0.133 & 0.018 & 5\\
1470 &  -0.104 & 0.023 & 5\\
1471 &  -0.068 & 0.019 & 4\\ 
1472 &  -0.075 & 0.014 & 4\\
1473 &  -0.094 & 0.022 & 4\\
1474 &  -0.107 & 0.209 & 2\\
1477 &   0.143 & 0.180 & 4\\
1478 &  -0.099 & 0.009 & 4\\ 
1481 &  -0.420 & 0.003 & 3\\
1484 &  -0.089 & 0.125 & 5\\   
1485 &  -0.063 & 0.207 & 5\\   
1486 &  -0.034 & 0.034 & 4\\   
1490 &   0.094 & 0.105 & 3\\   
1491 &  -0.099 & 0.020 & 4\\   
1492 &   0.016 & 0.014 & 3\\   
1499 &  -0.073 & 0.011 & 3\\   
1504 &  -0.038 & 0.038 & 4\\   
1507 &  -0.094 & 0.026 & 4\\   
1510 &  -0.002 & 0.037 & 4\\   
1513 &  -0.102 & 0.009 & 4\\  
1526 &  -0.263 & 0.139 & 2\\
\hline
\end{tabular}
\end{table}

\begin{figure}
\begin{center}
\leavevmode  
\psfig{file=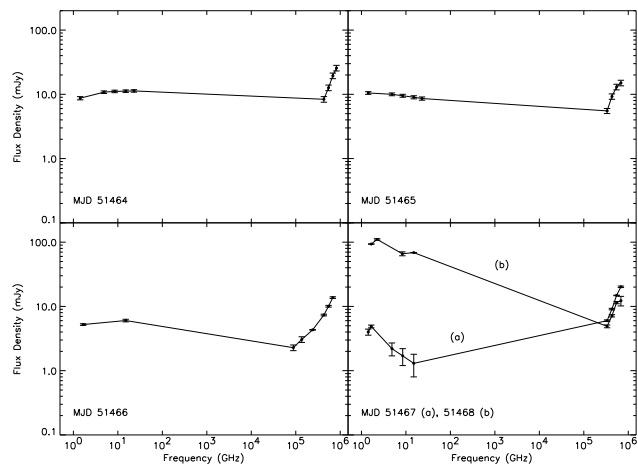,height=8.5cm,angle=90}  
\caption{The broad band spectrum for the first four epochs of radio observations, showing the flat spectrum in the first two; the spectrum then becomes optically thin in the fourth epoch. Optical and infrared data accumulated from IAU circulars have also been included (see text for references) and show that it is not unfeasible that the flat synchrotron spectrum of (at least) the first two epochs extends into the near-infrared. The optical data does not appear to fit the synchrotron spectrum (unlike that of XTE J1118+480, Markoff et al. 2001) and instead appears to be dominated by thermal irradiated-disc emission (Hynes et al., 2001) }.
\label{spectrum}
\end{center}
\end{figure}

In order to study the radio lightcurves in more detail they have been plotted again in Fig.~\ref{radio-lightcurve} on an expanded scale and using different symbols to indicate the different frequencies. It is now clear that the outburst did not just consist of a single flare -- indeed, our data show that at least five flares probably took place. The estimated time of each of these flares (and also the first hard X-ray flare) is indicated by arrows.

The bottom panel of Fig.~\ref{radio-lightcurve} shows the evolution of the spectral index ($S_{\nu} \propto \nu^{\alpha}$ where $S$ is the flux density, $\nu$ is the frequency and $\alpha$ is the spectral index); we also list the values of $\alpha$ in Table~\ref{alpha}, indicating the number of frequencies that were available for its calculation from a power law fit. Note that the radio fluxes have been daily averaged in order to determine the spectral indices -- unfortunately this has the effect of increasing the uncertainty in the spectral indices. The radio emission throughout most of the outburst displays a spectrum which is indicative of a combination of optically thick and optically thin regions -- the spectrum generally remains flatter than $\alpha = -0.2$. This could be explained by a new ejection event beginning before the previous plasmon has expanded to the point at which the optical depth to self-absorption, $\tau, < 1$.

However, it is also clear to see that the spectral index becomes positive, i.e. the spectrum inverts, on a number of occasions. These tend to take place just prior to a mass ejection. It would appear that the source repeatedly ejects an optically thick plasmon which later expands and gradually becomes more optically thin to successively decreasing frequencies. This is not unusual and has been seen in other sources, such as A0620$-$00 (Kuulkers et al. 1999).

The first of these spectral inversions is during the initial observations, at which time the X-ray spectrum is still dominated by the hard flux. We note that a continuous radio jet with a flat ($\alpha\sim 0$) spectrum (due to partial synchrotron self-absorption) is a ubiquitous feature of low/hard state sources (see e.g. Fender 2001, Brocksopp et al. 2001); it appears that in the case of XTE J1859+226 also, the production of harder X-rays is closely linked with the generation of a jet, the spectrum of which shows evidence for some self-asorption. This initial hard state period is not necessarily a typical feature of a soft X-ray transient outburst and we address this further in Section 4.

The radio spectrum during the first four epochs has been plotted in Fig.~\ref{spectrum} in order to emphasize the shape of the spectrum -- it is flat in the first two epochs, becoming optically thin in the fourth epoch. We also include optical and infrared data from the literature (Norton et al. 1999, Chaty et al. 1999) -- these data have been de-reddened using E(B$-$V)=0.58 (Hynes et al. 1999) and the frequency extrapolation of Cardelli et al. (1989). It is clear from these plots that the optical emission is thermal as we would expect -- the synchrotron spectrum cannot dominate shortwards of the near-infrared. However, we note that it is not impossible that the flat spectrum extends into the near-infrared, particularly if there was $K$ band variability during the decline as hinted at in Fig. 1 of Chaty et al. (2001) -- if the infrared emission was disc-dominated then we would expect a non-variable lightcurve, similar to that of the optical. We also note that the UKIRT points in Fig. 4 of Hynes et al. (2001) do not fit the X-ray irradiated disc model particularly well; the authors agree that an extension of the synchrotron spectrum to infrared frequencies is a possibility. 

We have fitted the spectrum in the first two epochs with a straight line for different frequency ranges -- 1.42 GHz up to each of $U$, $B$, $V$, $R$ and $I$. For both of these epochs the best fit line ($\chi^{2}_{\nu}=1.6$) has $\alpha \sim0$ for the 1.42 GHz -- $R$ band range, suggesting that a flat synchrotron spectrum up to the $R$ band is entirely possible, as found for a number of other jet-emitting X-ray binaries -- such as Cyg X-3, GRS 1915+105 and possibly GS 1354$-$64 (see e.g Fender 2001). However, in the case of XTE J1859+226 the optical emission is probably more consistent with an irradiated disc spectrum -- see Hynes et al. (2001). We note that the synchrotron spectrum {\em may} extend further -- however it becomes dominated by the X-ray irradiation and can no longer be observed. The power requirements for a jet emitting synchrotron radiation up to infrared wavelengths and the subsequent implications on current accretion models are explained in Fender (2001); see also Spencer (1999) and references therein.

\begin{figure*}
\begin{center}
\leavevmode  
\psfig{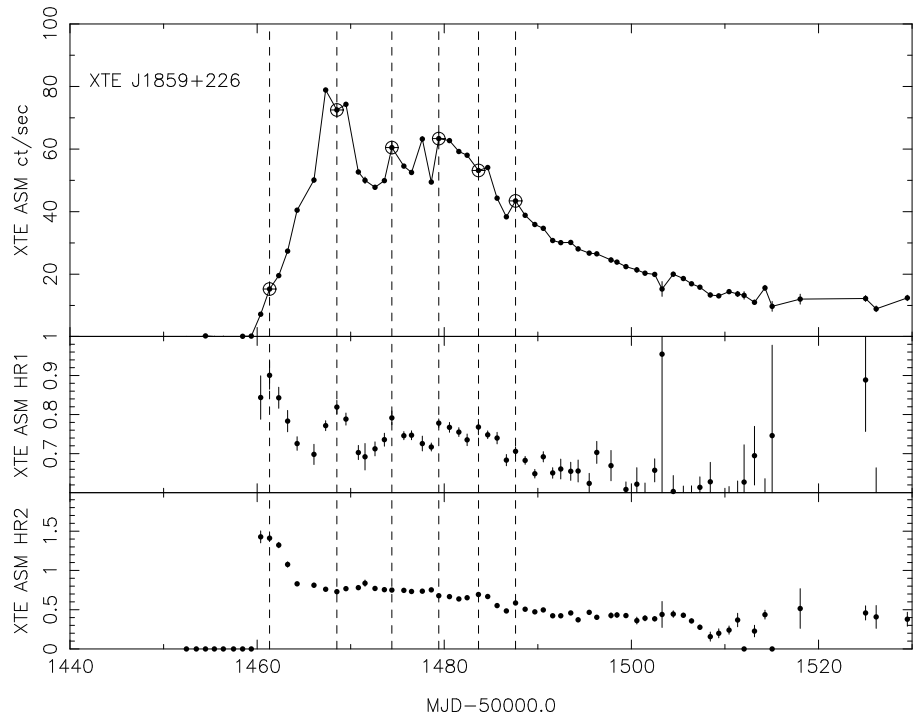}  
\caption{Daily averaged ASM data plotted with the hardness ratios (HR1 and HR2, see text). Dashed lines and circled points correspond to the estimated times of the initial hard X-ray flare and radio ejections and clearly correlate with periods of X-ray spectral hardening.}
\label{hardness}
\end{center}
\end{figure*}

\begin{figure*}
\begin{center}
\leavevmode  
\psfig{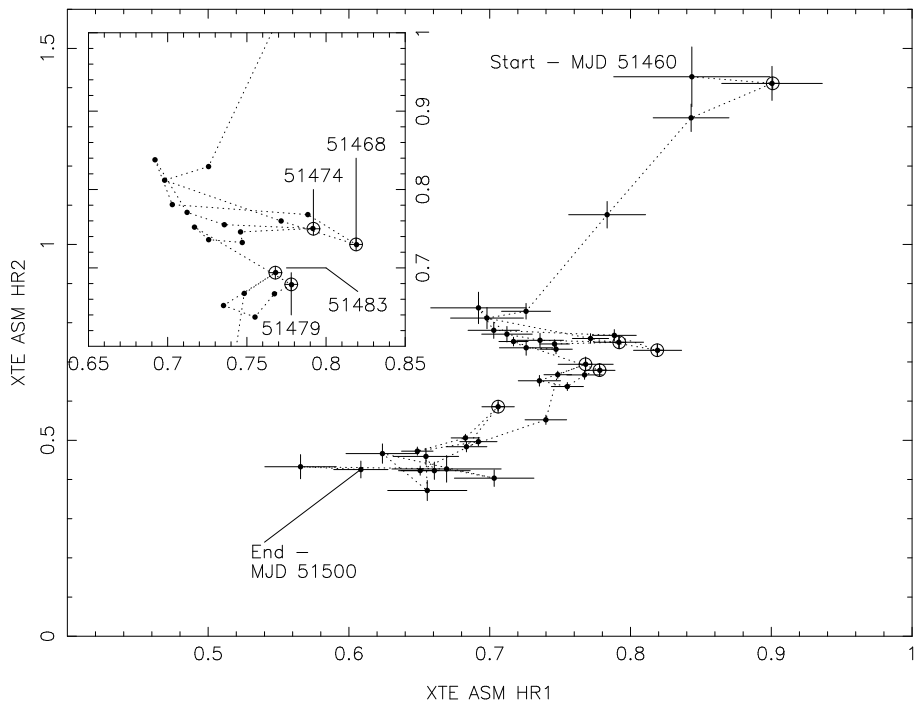}  
\caption{X-ray colour-colour diagram showing the evolution of the spectrum during the outburst -- again, the daily averaged data from Fig.~\ref{hardness} was used. Spectral hardening takes place towards the top right corner, spectral softening towards the bottom left corner. The inset shows the central regions of the plot in more detail. The circled points again indicate the times of the hard X-ray peak and radio ejections and again correspond to times of spectral hardening.}
\label{colour}
\end{center}
\end{figure*}

\subsection{X-ray}

The soft X-ray lightcurve is plotted again in Fig.~\ref{hardness} along with the hardness ratios (HR1 is the ratio of counts in the 3--5 keV band to the 1.5--3 keV band; HR2 compares counts in the 5--12 keV and 3--5 keV bands). It is now easy to see that the source was initally in a hard state and then softened with the soft X-ray peak. Dashed lines corresponding to the arrows marked on Fig.~\ref{radio-lightcurve} have been added to the plot. It can be seen that, while generally the source softens during the decline of the outburst, there are a number of (quasi-periodic) events during which the spectrum hardens and that these events are very well correlated with the radio events. 

This pattern of behaviour is emphasized in Fig.~\ref{colour} -- an X-ray colour-colour diagram showing the overall evolution towards a softer X-ray spectrum, superimposed on which are temporary hardenings associated with the radio ejections. The circled points correspond to the radio events as indicated in Fig.~\ref{hardness} and again show clearly that the radio ejections are accompanied by spectral hardening (or perhaps vice versa?). It can also be seen that, since the variability of HR2 is minimal, these hardenings may be the result of a decrease in the 1.5--3 keV band flux at the time of an ejection event -- unfortunately an investigation of the lightcurve at each ASM band was inconclusive. We note the possibility that our radio monitoring may have missed a radio flare coincident with the first time of X-ray hardening (i.e. the first and major hard X-ray flare), although comparisons with other sources shows that this need not be the case (see Section 4).

We note that the general softening of the X-ray source during the decline of the outburst is not considered a typical characteristic. Of the soft X-ray transients observed by {\it RXTE} it appears that only XTE J2012+381 and 4U 1630$-$47 share this property (and also the hard state transients XTE J1118+480 and GS 1354$-$64) -- indeed one might expect that the source should harden as the luminosity decays. Low/hard state behaviour during the decay has been observed in XTE J1550$-$564 (Homan et al. 2001) and GRO J1655$-$40 (M\'endez et al. 1998) and given the very similar natures of these two sources and XTE J1859+226 (see Section 4) it is perhaps surprising to find such a noticeable difference. 

\subsection{Radio/X-ray correlations}

In addition to the radio/spectral hardening correlations seen above, Fig.~\ref{lightcurve} indicates correlated behaviour between the broad-band X-ray and radio lightcurves and this would be expected from study of other X-ray binaries in which correlated behaviour has been observed (see Section 4).

\begin{figure}
\begin{center}
\leavevmode  
\psfig{file=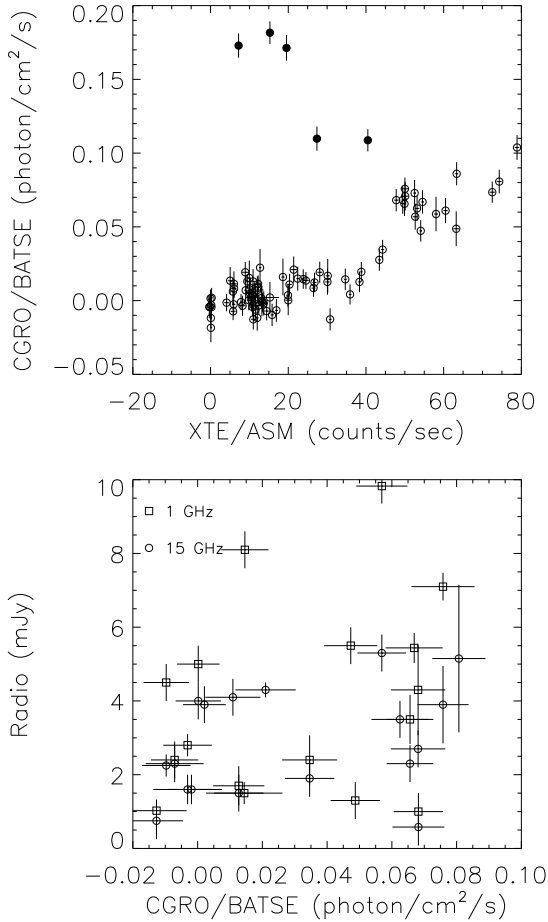,height=14cm,angle=0}  
\caption{Daily averaged flux-flux plots. Top: The hard X-rays plotted against the soft X-rays shows an anti-correlation during the first flare (solid symbols) and a strong ($r\sim0.83$) correlation following the first flare (hollow symbols). Bottom: The radio flux densities plotted against the hard X-ray data, using data from after the first flare only. A reasonable correlation ($R\sim 0.68$) is observed. See text and Table~\ref{spearman} for details.}
\label{correlate}
\end{center}
\end{figure}

Spearman rank correlation coefficients have been calculated for pairs of datasets, using all daily averaged data points, and are shown in Table~\ref{spearman}. They suggest significant overall correlations between the soft X-rays and the radio (but see below), between the hard and soft X-rays and between the two radio bands (we note that time lags have not been taken into account). Furthermore, these values can be increased by taking subsets of the data -- in particular the hard/soft X-ray correlation has coefficient 0.83 if the initial hard X-ray flare is ignored (top plot in Fig.~\ref{correlate}, ignoring the filled symbols); likewise the hard X-ray and 15 GHz emission have a correlation coefficient of 0.68 ($>$ 99\% confidence) if the first bright radio flare is ignored (bottom plot in Fig.~\ref{correlate}).

Thus it appears that the first and major radio ejection is correlated with the {\em soft} X-ray emission only. Following this flare however, the correlation between the {\em hard} X-rays and radio emission is more significant -- given the spectral hardening at the time of each radio event this is not surprising. Indeed, ignoring the first radio flare brings the soft X-ray/radio correlation coefficient down to 0.11 -- this suggests that the clear soft X-ray/radio relationship which exists during the first radio flare becomes much more complex during the subsequent series of ejections. During this latter period it appears that the radio behaviour is linked more directly with the hard X-rays. Since spectral hardening takes place during the soft X-ray peak it may actually be the case that the accretion disc does {\em not} dominate the soft X-ray emission at this time. Alternatively it is possible that the correlation between the hard X-rays and the radio emission is strong also during the initial flares -- but that the radio event is delayed by a few days. The implications of such correlations are discussed in Section 4.2.

\begin{table}
\center
\caption{Spearman rank correlation coefficients ($r$) obtained for various pairs of datasets. We use Student's $t$ test to confirm that all correlations have a 99\% confidence limit (with the exception of the BATSE/radio values which are 98\% and 95\% for the 15 and 1 GHz respectively). Note that these values can be improved significantly by considering subsets of the data -- see text for details.}
\label{spearman}
\begin{tabular}{cccc}
\hline
&&$r$&No. of points\\ 
\hline
ASM&1 GHz&0.59 &23\\
ASM&15 GHz&0.58 &22\\
BATSE&1 GHz&0.46 &24\\
BATSE&15 GHz&0.50&23\\
ASM&BATSE&0.55&90\\
1 GHz&15 GHz&0.90&17\\
\hline
\end{tabular}
\end{table}

\subsection{Energetics}

It is extremely unusual to observe the rise to a radio outburst and we have been very fortunate in obtaining excellent MERLIN coverage of the major radio flare of XTE J1859+226 in 1999 (Fig.~\ref{merlin}). As we can therefore estimate the rise time of the event and if we assume that the radio source is in a state of approximate equipartition then it is possible to determine the minimum energy ($W_{\mbox{min}}$) associated with an ejection event (see e.g. Longair 1994):

\begin{equation}
W_{\mbox{min}}\approx 3.0\times10^6\eta^{4/7}\bigg(\frac{V}{\mbox{m}^3}\bigg)^{3/7}\bigg(\frac{\nu}{\mbox{Hz}}\bigg)^{2/7}\bigg(\frac{L_{\nu}}{\mbox{W\,Hz}^{-1}}\bigg)^{4/7}\,\,\mbox{J}
\end{equation}

The flare reached $S_o \sim 110$ mJy at $\nu_o = 1.66$ GHz; the spectral index, $\alpha$ (where the flux density $S_{\nu}\propto \nu^{\alpha}$) at the peak was $-0.09$ for the range $1.42<\nu<23$ GHz. Thus the luminosity ($L_{\nu}=4\pi d^2S_o\nu_o^{-\alpha}\int^{\nu^2}_{\nu_1}\nu^{\alpha}\mbox{d}\nu \,\,\mbox{erg/s}$) of the source was $2.42\big(\frac{d}{\mbox{kpc}}\big)^2\times 10^{30}$ erg/s. The rise time of the event was $\sim 6$ hours; we can therefore assume that the (spherical) ejection had a volume of $V\sim 10^{45} \mbox{cm}^3$. Finally we assume that the relativistic proton energies are negligable compared with those of the electrons (Fig.~\ref{merlin}) to give $\eta=1$. Therefore the minimum energy required to produce the major radio ejection was:

\noindent $W_{\mbox{min}}\sim 2.4\big(\frac{d}{\mbox{kpc}}\big)^{8/7}\times 10^{33} \mbox{J} = 2.4\big(\frac{d}{\mbox{kpc}}\big)^{8/7}\times 10^{40} \mbox{erg}$.\\
By dividing this by the rise time we obtain the minimum jet power:

\noindent $P_{\mbox{min}}=1.1\big(\frac{d}{\mbox{kpc}}\big)^{8/7}\times 10^{29} \mbox{J/s} = 1.1\big(\frac{d}{\mbox{kpc}}\big)^{8/7}\times 10^{36} \mbox{erg/s}$.

The low extinction of XTE J1859+226 ($E(B-V)\sim 0.58$) might suggest that the distance to the source is fairly small. Note that this is almost certainly a conservative lower limit to the power going into the jet production, as we do not measure the high-frequency extent of the synchrotron spectrum, nor do we take account of energy associated with bulk relativistic motion.  Nonetheless, our estimate shows that the jet of  XTE J1859+226 still uses an extremely significant fraction of the accretion power, although this problem can be somewhat alleviated by the internal shock model of Kaiser, Sunyaev \& Spruit (2000).

\section{Discussion and Comparison with other sources}

\subsection{The initial low/hard state / hard X-ray flare}

Given the ultrasoft spectrum of most of these soft X-ray transient events it might be expected that the initial rise to outburst is dominated by the soft component. Indeed, this has been seen in the 1996 outburst of GRO J1655$-$40 when there was a $\sim30$ day delay before the hard X-ray flux increased. With such emphasis in the literature on the `canonical' FRED (Fast Rise Exponential Decay) outburst (e.g. Chen, Shrader \& Livio 1997 and references therein) there is little in the literature to dispute that this is normal.

However in the case of the 1994 outburst of GRO J1655$-$40, XTE J1550$-$564, 4U 1630$-$47 (e.g. 1998 and 1999 events) and now XTE J1859+226 it has been the other way round. Instead of a sudden soft X-ray rise there has been a sudden hard X-ray peak, accompanied by a gradual soft X-ray increase. In the case of XTE J1859+226 the soft X-ray peak accompanied the decay of the hard X-rays. A second and more luminous hard X-ray peak accompanied the soft X-ray outburst in XTE J1550$-$564 and there was no soft X-ray information in the case of GRO J1655$-$40.

This hard state behaviour prior to a soft X-ray transient event is not accounted for in FRED models and so is perhaps unexpected. Maybe given that a number of sources outburst from quiescence into the hard state and do not manage to soften throughout their outbursts we should not find this result surprising. Likewise Cyg X-1, LMC X-3 and GX 339$-$4 have all entered the high/soft state from the low/hard state (e.g. Brocksopp et al. 1999, Wilms et al. 2001, Maejima et al. 1984). Unfortunately there is no spectral information for the radio events in XTE J1550$-$564 during the initial hard state period (although we note that a flat radio spectrum was present during a hard state period of a subsequent outburst, Corbel et al. 2001) -- the constant flux density would suggest that a flat spectrum jet was present in the early phases of that outburst as well as that of XTE J1859+226. Again, no early radio data are available for GRO J1655$-$40.

With our current ability to detect outbursting sources very quickly, both initally at X-ray wavelengths and subsequently with ground-based telescopes, we now have the means to study a source's rise to outburst. If we wish to understand the mechanisms behind a soft X-ray transient outburst then we are most likely to obtain our information from the inital low/hard state period (particularly if this is a more common feature than we realise -- see Section 4.5) -- it may even be that this is the time when the fuel to power the outburst is accumulating. Observing the decay is less beneficial as by this later stage the source is just cooling gradually. Furthermore, we have {\em already} studied this latter stage for many sources and still we do not know what drives the outburst. More importantly, we still do not know how the outbursts of the disc, corona and jet relate to one another -- an alternative method appears to be necessary. Fortunately there is a large amount of low/hard state data available to study; in particular the low/hard to high/soft transitions (and vice versa) of persistent sources such as Cyg X-1, GX 339$-$4 and LMC X-3 can provide valuable analogues of a soft X-ray transient outburst.

\begin{table*}
\center
\caption{A list of the known soft X-ray transients (top) and the hard state transients (bottom) with a summary of their X-ray and radio observations at the time of outburst. Key to symbols: $\bullet$ -- yes, x -- no, $\bullet$? -- probably, ? -- insufficient data to be certain either way, $\ast$ -- the hard state sources were found to emit flat spectrum radio jets. Analysis of this table shows that there is no source which definitely did not emit multiple radio ejections and no source that definitely did not spend some time (i.e. $\sim$ few days) in the hard state prior to the transition to a softer state.}
\label{selection}
\begin{tabular}{lccccccccccl}
\hline
Source&Year&\multicolumn{3}{c}{Observed at all?}&\multicolumn{3}{c}{Observed Initial Rise?}&$\ge2$ Radio&Initial&Hard&Refs.\\
\cline{3--4}
\cline{4--5}
\cline{5--5}
\cline{6--7}
\cline{7--8}
\cline{8--9}
&&Hard&Soft&Radio&Hard&Soft&Radio&Ejection&Hard&State&\\
&&X-rays&X-rays&&X-rays&X-rays&&Events?&State?&Only?&\\
\hline
GS 1354$-$64&1966/1987&$\bullet$&$\bullet$&x&x&x&x&?&?&x&1\\
4U 1543$-$475&1971/1983&$\bullet$/x&$\bullet$&x&x&x/$\bullet$&x&?&?/$\bullet$?&x&2,3,4\\
A1524$-$617&1974&$\bullet$&$\bullet$&x&x&$\bullet$&x&?&?&x&5\\
A0620$-$00&1975&$\bullet$&$\bullet$&$\bullet$&$\bullet$&x&x&$\bullet$&$\bullet$?&x&6,7\\
H 1705$-$250&1977&$\bullet$&$\bullet$&x&x&$\bullet$&x&?&?&x&8,9\\
H 1741$-$322&1977&$\bullet$&$\bullet$&x&x&x&x&?&?&x&10\\
EXO 1846$-$031&1985&$\bullet$&$\bullet$&x&x&x&x&?&?&x&11\\
GS 2000+251&1988&x&$\bullet$&$\bullet$&x&$\bullet$&x&?&?&x&10\\
GRS 1124$-$684&1991&$\bullet$&$\bullet$&$\bullet$&$\bullet$&$\bullet$&x&$\bullet$&$\bullet$?&x&12,13\\
GRS 1009$-$45&1993&$\bullet$&$\bullet$&x&$\bullet$&$\bullet$?&x&?&$\bullet$&x&14\\
GRO J1655$-$40&1994&$\bullet$&x&$\bullet$&$\bullet$&x&x&$\bullet$&$\bullet$&x&15\\
GRS 1730$-$312&1994&$\bullet$&$\bullet$&x&$\bullet$&?&x&?&$\bullet$?&?&16\\
GRS 1739$-$278&1996&$\bullet$&$\bullet$&$\bullet$&$\bullet$&$\bullet$&x&?&$\bullet$&x&17,18,19\\
4U 1630$-$47&e.g. 1994&x&$\bullet$&x&x&x&x&?&?&x&20\\
4U 1630$-$47&1998&$\bullet$&$\bullet$&$\bullet$&$\bullet$&$\bullet$&x&?&$\bullet$&x&21\\
XTE J1748$-$288&1998&$\bullet$&$\bullet$&$\bullet$&$\bullet$&$\bullet$&x&?&$\bullet$&x&22,23,24\\
XTE J1550$-$564&1998&$\bullet$&$\bullet$&$\bullet$&$\bullet$&$\bullet$&$\bullet$&$\bullet$&$\bullet$&x&25\\
XTE J2012+381&1998&$\bullet$&$\bullet$&$\bullet$&$\bullet$&$\bullet$&x&?&$\bullet$&x&26\\
XTE J1859+226&1999&$\bullet$&$\bullet$&$\bullet$&$\bullet$&$\bullet$&$\bullet$&$\bullet$&$\bullet$&x&27\\
XTE J1819$-$254&1999&$\bullet$&$\bullet$&$\bullet$&$\bullet$&$\bullet$&x&$\bullet$?&$\bullet$?&x&28\\
\hline
GS 2023+338&1989&$\bullet$&$\bullet$&$\bullet$&x&x&x&$\ast$&$\bullet$&$\bullet$&10\\
A1524$-$617&1990&$\bullet$&x&x&x&x&x&?&?&$\bullet$?&29\\
4U 1543$-$475&1990&$\bullet$&x&x&$\bullet$&x&x&?&$\bullet$?&$\bullet$?&30\\
GRO J0422+32&1992&$\bullet$&x&$\bullet$&$\bullet$&x&x&$\ast$&$\bullet$&$\bullet$&31,32\\
GRO J1719$-$24&1993&$\bullet$&x&$\bullet$&$\bullet$&x&x&$\ast$&$\bullet$&$\bullet$&33,34\\
GS 1354$-$64&1997&$\bullet$&$\bullet$&$\bullet$&$\bullet$&$\bullet$&x&$\ast$&$\bullet$&$\bullet$&35\\
XTE J1118+480&2000&$\bullet$&$\bullet$&$\bullet$&$\bullet$&$\bullet$&x&$\ast$&$\bullet$&$\bullet$&35,36\\
\hline
\end{tabular}
\begin{tabular}{l}
\\
1. Lewin et al. (1968) 2. Li et al. (1976) 3. Matilsky et al. (1972) 4. Kitamoto et al. (1975) 5. Kaluzienski et al. (1975)\\
6. Ricketts et al. (1975) 7. Kuulkers et al. (1999) 8. Watson et al. (1978) 9. Wilson \& Rothschild (1983) 10. Tanaka \& Lewin (1995)\\
11. Parmer et al. (1993) 12. Ball et al. (1995) 13. Ebisawa et al. (1994) 14. Lapshov et al. (1993) 15. Harmon et al. (1995)\\
16. Churazov et al. (1994) 17. Vargas et al. (1997) 18. Borozdin et al. (1996) 19. Hjellming et al. (1996a)\\
20. Parmar et al. (1997) 21. Hjellming et al. (1999) 22. Smith et al. (1998) 23. Harmon et al. (1998) 24. Hjellming et al. (1998)\\
25. Hannikainen et al. (2001) 26. Vasiliev et al. (2000) 27. This paper 28. Hjellming et al. (2000)\\
29. Barret et al. (1992) 30. Harmon et al. (1992b) 31. Shrader et al. (1994) 32. Callanan et al. (1995) 33. Revnivtsev et al. (1998)\\
34. Hjellming et al. (1996b) 35. Brocksopp et al. (2001) 36. Wilson \& McCollough (2000)\\
\hline
\end{tabular}
\end{table*}

\subsection{Multiple radio ejections and X-ray/radio correlations}

Radio monitoring of the persistent X-ray binaries such as GX 339$-$4 has shown that, whereas the low/hard state is characterised by a flat spectrum continuous jet, when the source makes the transition to the high/soft state the jet is quenched -- with its flux density below the sensitivity limit of radio telescopes (e.g. Fender 2001 and references within). As the source makes the transition, however, a plasmon ejection may take place (Corbel et al. 2000). In such cases the soft state can be temporarily associated with (fading) radio emission from ejecta which are however by this time physically decoupled from the accretion process (Fender \& Kuulkers 2001).

This appears also to be the case for the soft X-ray transients. As they rise above quiescence and become softer, the radio emission indicates ejection of material; it then expands, becomes optically thin and decays. In a number of sources this has been repeated on timescales of days to months -- in sources A0620$-$00 and GRO J1655$-$40 a number of ejections took place and a second ejection is also apparent in the lightcurves of GRS 1124$-$684 and possibly XTE J1550$-$564 (see e.g. Kuulkers et al. 1999, Harmon et al. 1995 and Hannikainen et al. 2001). GRO J1655$-$40 was particularly comparable with XTE J1859+226 as both sources (and also GRS 1915+105, see below) showed a sequence of decaying amplitude radio flares accompanied by X-ray spectral hardening -- although it is interesting that the intervals between the ejections of XTE J1859+226 were much smaller ($\sim$ days, compared with tens of days) and decreased with time since the initial outburst (the intervals between the ejections of GRO J1655$-$40 increased). 

Although we might have assumed that ejection events are initiated by increased mass flow through the accretion disc it is interesting to see that the radio ejections of A0620$-$00 and GRS 1124$-$684 are not well-correlated with the disc behaviour. Similarly in the case of XTE J1859+226 the lightcurves are not well-correlated following the initial simultaneous flare. On the contrary, peaks in the X-ray power-law emission of GRS 1124$-$684 and GRO J1655$-$40 (and XTE J1859+226 after MJD 51470, 1999 Oct 18) {\em are} correlated with the radio ejections -- it may be the case that the first radio flare of XTE J1859+226 is actually triggered by the BATSE flare five days previously. Such a radio lag is a typical feature of these ejections (e.g. Kuulkers et al. 1999).

Correlated behaviour between X-ray and radio emission has also been observed in other black hole X-ray binaries (see e.g. Fender 2001, and references therein); Cyg X-1 and GX 339$-$4 have both shown correlated X-ray/radio flares whilst in the low/hard state, and are well known for the dramatic quenching of both radio and hard X-ray emission on transition to the high/soft state. GRS 1915+105 (see below) and Cyg X-3 have also exhibited dramatic flares simultaneously in X-ray and radio emission.

Previous authors have commented on the importance of understanding the mechanisms by which material is passed through the accretion disc and then ejected into jets -- e.g. Fender \& Kuulkers (2001) show that there is a correlation between the peak X-ray and radio fluxes during transient events, thus implying a relationship between the peak accretion rate and the number of synchrotron-emitting electrons ejected. This is clearly a non-straightforward process as seen in Fig.~\ref{lightcurve}, with the series of flares taking place only {\em quasi}-simultaneously at all three wavebands. Our observations might suggest that the coupling between the jet and the corona is even more significant than the coupling between the jet and the disc -- the correlations between the radio and hard X-rays tend to be more convincing. This is perhaps not surprising given that both synchrotron and inverse Compton sources require a (the same?) supply of relativistic electrons near the base of the jet. We discuss this further in a future paper.

\subsection{GRS 1915+105}

While GRS 1915+105 is perhaps not a good source with which to draw comparisons on account of its complexities, it has indeed displayed very similar behaviour to XTE J1859+226. The radio lightcurve can be seen in Fig. 1 of Fender et al. (1999), along with plots showing the accompanying soft X-ray and radio spectral index behaviour. While the luminosity and variability of GRS 1915+105 are considerably more dramatic than XTE J1859+226, it is clear that there is a period during which the radio spectrum is flat (possibly out to infrared frequencies -- see Fender 2001) and the X-ray spectrum is hard -- comparable with hard state behaviour. The source then enters a series of correlated outbursts, each associated with an ejection event and again decreasing in amplitude and possibly also in time interval since the previous ejection -- very similar to a scaled-up version of XTE J1859+226. We note that in the case of GRS 1915+105, these multiple ejections of decreasing strength were directly resolved and tracked as they moved relativistically away from the system (Fender  et al. 1999).

\subsection{Selection effects ?}

It is tempting to designate sources such as XTE J1859+226, XTE J1550$-$564, GRO J1655$-$40 and possibly also A0620$-$00 as `strange'; they are the only sources to have been observed to emit a number of radio ejections or to have remained in a low/hard state for a few days prior to reaching the full ultrasoft outbursts for which they are so well-known. However it is also important not to complicate our knowledge of X-ray binaries with more and more sub-classification.

For this reason we have searched the literature for the X-ray and radio properties of all black hole X-ray transients, both ultrasoft and hard state sources -- the results are in Table~\ref{selection}, with soft X-ray transients at the top and hard state transients below. We indicate which sources were observed at hard X-ray ($>15$ keV), soft X-ray ($<15$ keV) and radio wavelengths, in which cases the initial rise of the outburst was monitored and whether multiple radio ejections or initial low/hard state behaviour were observed.

It is now possible to see that XTE J1859+226, XTE J1550$-$564, GRO J1655$-$40 and A0620$-$00 may not be strange after all. In order to observe multiple radio ejections, continuous radio monitoring is required -- it is common for the rise and decay of an ejection event to take place within a day (see Fig.~\ref{lightcurve}) -- and this has not always been possible. Likewise, in order to observe an initial hard state the outburst must be observed immediately as it leaves quiescence -- this has been achieved very rarely in the past. {\em There is no source which definitely did not emit multiple radio ejections and no source that definitely did not spend some time (i.e. $\sim$ few days) in the hard state prior to the transition to a softer state}. The one exception is the 1996 outburst of GRO J1655$-$40 (Hynes et al. 1998) --  in this case it is possible that the 1994 outburst had diminished the supply of relativistic electrons required for hard X-ray and radio emission and it was not until increased mass flow through the disc had replenished this supply that the jet and corona emission could begin. If this was the case then we must ask how the electron supply is produced prior to the initially hard events.

\section{Conclusions}

We have observed XTE J1859+226 at radio and X-ray wavelengths during its 1999 outburst. The results show that this soft X-ray transient spent a few days in the low/hard state before increasing to its maximum and we suggest that this may be a typical feature of soft X-ray transient outbursts, thus providing a means by which we can study the outburst mechanisms. As the outburst continued we found that a series of radio ejections took place, simultaneously with spectral hardening and apparently correlated with the hard X-ray lightcurve, thus emphasizing the strength of the jet/corona connection. Again, it appears that a series of radio ejections correlated with the corona emission may be typical features of these systems.

\section*{acknowledgements}

MERLIN is a national facility operated by the University of Manchester on behalf of PPARC. We also thank the staff at MRAO for the maintenance and operation of the RT, which is supported by PPARC. The Green Bank Interferometer was operated by the National Radio Astronomy Observatory for the U.S. Naval Observatory and the Naval Research laboratory during the time-period of these observations. The {\it RXTE} ASM data used in this paper was obtained from the public ASM database.

CB acknowledges a PPARC studentship and Open University travel budget at the time of the outburst, as well a PPARC grant and STARLINK facilities at JMU; CJF acknowledges a PPARC research stipend. S.A.T. thanks the Russian Foundation of Base Researches (RFBR) for support through grant N98-02-17577. Final thanks to Bryan Anderson and Graham Smith for useful comments.

\end{document}